\documentclass[usegraphicx,referee]{mn2e}
\usepackage{times}
\newlength{\colwidth}
\begin{document}

\title[Displacement of the Sun from the Galactic Plane]
      {Displacement of the Sun from the Galactic Plane}
\author[Y. C. Joshi]{Y. C. Joshi\thanks{E-mail: y.joshi@qub.ac.uk}\\
Astrophysics Research Centre, School of Mathematics and Physics, Queen's University
Belfast, Belfast BT7 1NN, UK\\
}
\date{Accepted 2007 April 5;
Received 2006 July 5;
}  

\pagerange{\pageref{firstpage}--\pageref{lastpage}} \pubyear{2007}

\maketitle

\label{firstpage}

\begin{abstract} 
We have carried out a comparative statistical study for the displacement of the Sun
from the Galactic plane ($z_\odot$) following three different methods. The
study has been done using a sample of 537 young open clusters (YOCs) with
$\log({\rm Age}) < 8.5$ lying within a heliocentric distance of 4 kpc and 2030 OB stars
observed up to a distance of 1200 pc, all of them have distance information.
We decompose the Gould Belt's member in a statistical sense before investigating
the variation in the $z_\odot$ estimation with different upper cut-off
limits in the heliocentric distance and distance perpendicular to the Galactic
plane. We found $z_\odot$ varies in a range of $\sim 13 - 20$ pc from the analysis
of YOCs and $\sim 6 - 18$ pc from the OB stars. A significant scatter in the
$z_\odot$ obtained due to different cut-off values is noticed for the OB stars
although no such deviation is seen for the YOCs. We also determined scale heights
of $56.9^{+3.8}_{-3.4}$ and $61.4^{+2.7}_{-2.4}$ pc for the distribution of YOCs
and OB stars respectively.
\end{abstract}


\section{Introduction}
It has long been recognized that the Sun is not located precisely in the mid-plane of the
Galactic disk defined by $b = 0^\circ$ but is displaced a few parsecs to the
North of Galactic plane (GP) (see Blitz \& Teuben 1996 for a review) and understanding
the exact value of $z_\odot$ is vital not only for the Galactic structure models but
also in describing the asymmetry in the density distribution of different kind of
stars in the north and south Galactic regions (Cohen 1995, M\'{e}ndez \& van Altena 1998,
Chen et al. 1999). Several independent studies in the past have been carried out to
estimate $z_\odot$ using different kind of astronomical objects, for example,
Gum, Kerr \& Westerhout (1960) concluded that $z_\odot = 4\pm 12$ pc from the neutral hydrogen
layer, Kraft \& Schmidt (1963) and Fernie (1968) used Cepheid variables to estimate
$z_\odot \sim 40$ pc while Stothers \& Frogel (1974) determined $z_\odot = 24\pm 3$ pc
from the B0-B5 stars within 200 pc from the Sun, all pointing to a broad range of
$z_\odot$. More recently various different methods have been employed to estimate $z_\odot$
e.g. Cepheid variables (Caldwell \& Coulson 1987), Optical star count technique
(Yamagata \& Yoshii 1992, Humphreys \& Larsan 1995, Chen et al. 2001),
Wolf-Rayet stars (Conti \& Vecca 1990), IR survey (Cohen 1995, Binney, Gerhard \& Spergel
1997, Hammersley et al. 1995) along with different simulations (Reed 1997, M\'{e}ndez \&
van Altena 1998) and models (Chen et al. 1999, Elias, Cabrera-Ca\~{n}o \& Alfaro 2006,
hereafter ECA06). Most of these studies constrained $z_\odot$ in the range of
15 to 30 pc in the north direction of the GP.

In recent years, the spatial distribution of open clusters (OCs) have been extensively used
to evaluate $z_\odot$ since continued compilation of new clusters has brought together more
extensive and accurate data than ever. Using the OCs as a diagnostic tool to determine
$z_\odot$, Janes \& Adler (1982) found $z_\odot$ = 75 pc for 114 clusters of age smaller
than $10^8$ yr while Lyng\.{a} (1982) determined $z_\odot \sim 20$ pc with 78 young
clusters up to 1000 pc. Pandey \& Mahra (1987) reported $z_\odot$ as 10 pc from the
photometric data of OCs within $|b| \le 10^\circ$ and Pandey, Bhatt \& Mahra (1988)
using a subsample of YOCs within 1500 pc obtained $z_\odot = 28 \pm 5$ pc. Most
recently, $z_\odot$ have been determined in three independent studies based on the
analysis of OCs. Considering about 600 OCs within $5^\circ$ of GP, we derived
$z_\odot = 22.8 \pm 3.3$ pc through the analysis of interstellar extinction in the
direction of the OCs (Joshi 2005, hereafter JOS05). Bonatto et al. (2006)
reported $z_\odot$ as 14.8 $\pm$ 2.4 pc using 645 OCs with age less than 200 Myrs
while Piskunov et al. (2006, hereafter PKSS06) estimated a value of 22 $\pm$ 4 pc
using a sample of
650 OCs which is complete up to about 850 pc from the Sun. On the other hand using
a few thousand OB stars within $10^\circ$ of the GP and 4 kpc from the Sun, Reed (1997)
approximately estimated the value as 10-12 pc while Ma\'{i}z-Apell\'{a}niz
(2001) determined this values as $24.2\pm2.1$ pc using a sample of about 3400 O-B5
stars obtained from the Hipparcos catalogue.

The large range of $z_\odot$ derived from these different methods could be possibly
caused by the selection of data of varying age, heliocentric distance $d$,
spectral type, etc. along with the method of the determination. The aim of the present paper is
therefore to study the variation in $z_\odot$ following different methods by
constraining different upper limits in $z$ and $d$ using a large
sample of OCs and OB stars. The paper is organized as follows.
First we detail the data used in this study in Sect. 2. In
Sect. 3, we examine the distribution of $z$ with the age of clusters while Sect. 4
deals their distribution with the different $z$ cut-off and $d$ cut-off in order to
determine $z_\odot$. The exponential decay of $z$
distribution of the OCs and OB stars and their variation over the Galactic longitude
are discussed in Sects. 5 and 6 respectively. Our results are summarized in Sect. 7.
\section {The Data}
We use two catalogues in this study. The OC catalogue is complied by
Dias et al.~(2002)\footnote{Updated information about the OCs is available in the
on-line data catalogue at the web site http://www.astro.iag.usp.br/$\sim$wilton/.}
which includes information available in the catalogues of the Lyng\.{a} (1987) as
well as WEBDA\footnote{http://obswww.unige.ch/webda} with the recent information
on proper motion, age, distance from the Sun, etc. The latest catalogue
(Version 2.7) that was updated in October 2006 gives physical parameters of
1759 OCs. Of these, 1013 OCs have distance information for which it is possible
to determine $z$ which is equivalent to $d \sin b$ where $b$ is the Galactic latitude. Out
of the 1013 OCs, age information is available for 874 OCs with ages ranging from 1 Myr to
about 10 Gyrs, although the majority of them are young clusters. Though the clusters are
observed up to a distance of about 15 kpc, it should be born in mind that
the cluster sample is not complete owing to large distance and/or low contrast of
many potential cluster candidates (Bonatto et al. 2006) and may be smaller by an
order of magnitude since a good fraction of clusters are difficult to observe at
shorter wavelengths due to large extinction near the GP (Lada \& Lada 2003, Chen,
Chen \& Shu 2004, PKSS06). When we plot cumulative distribution
of the clusters in our sample as a function of $d$ in Fig. 1,
we notice that the present cluster sample may not be complete beyond a distance of
about 1.7 kpc. A comprehensive discussion on the completeness of OCs has recently
been given by Bonatto et al. (2006) which along with PKSS06 puts the
total number of Galactic OCs in the order of $10^5$.
%-----------------------  Fig. 1 --------------------------------------------
\begin{figure*}
\includegraphics[height=12.0cm,width=16.0cm]{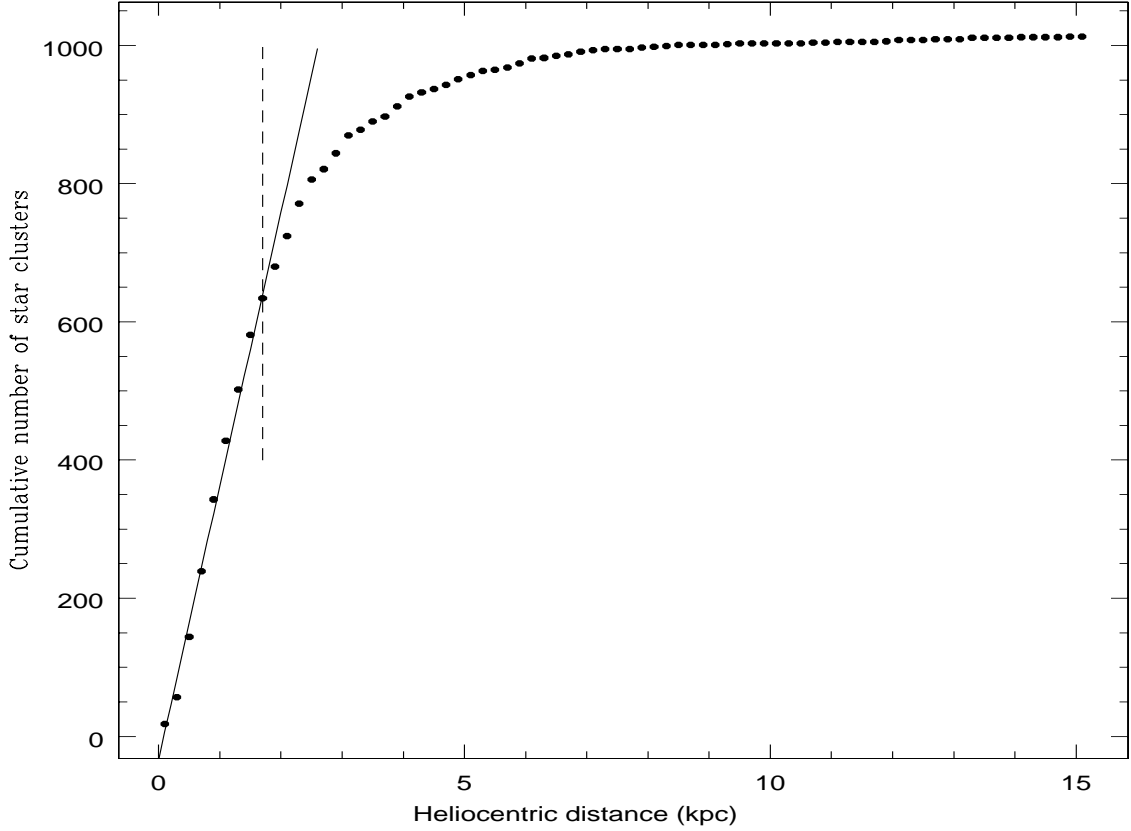}
\caption{A cumulative distribution diagram for the number of the open clusters with distance
from the Sun. The vertical dashed line indicates the completeness limit while continuous line
represents the least square fit in that region.}
\end{figure*}
%------------------------ End Fig. 1-----------------------------------------------

The other sample used in the present study is that of the OB stars 
taken from the catalogue of Reed (2006) which contains a total of 3457
spectroscopic observations for the 2397 nearby OB stars\footnote{For the detailed
 information about the data, the reader is referred to 
 http://othello.alma.edu/$\sim$reed/OBfiles.doc}. The distance of
OB stars are derived through their spectroscopic parallaxes. It 
is worth to note that the individual distance of OB stars may not be 
accurate (Reed 1997), nevertheless, a statistical study with significant 
number of OB stars can still be useful for the determination of $z_\odot$. Although,
 several studies on the determination of $z_\odot$ using OB stars have already
 been carried out on the basis of Hipparcos catalogue (Ma\'{i}z-Apell\'{a}niz 2001,
 ECA06 and references therein), however, it is noticed by some authors that the
 Hipparcos catalogue gives a reliable distance estimation within a distance of
 only 200-400 pc from the Sun (cf. Torra, Fern\'{a}ndez \& Figueras 2000). This is exactly the region
 where OB stars in the Gould Belt (hereafter GB) lie and this can cause an anomaly in the
 determination of $z_\odot$ if the stars belonging to the GB are
 not be separated from the data sample. Further Abt (2004) also noticed that
 classification of the stars in the Hipparcos catalogue is uncertain by about
  +/-1.2 subclass in the spectral classifications and about 10\% in the luminosity 
  classifications. In the present study we therefore preferred 
  Reed's catalogue of OB stars over the Hipparcos catalogue despite lesser in
  numbers but are reported up to a distance of about 1200 pc from the Sun and 
  $V \sim 10$ mag. The OB stars which have two different distances
  in the catalogue are assigned the mean distance provided they do not differ by more
  than 100 pc, otherwise we remove them from our analysis.
  If there are more than two distances available for any OB star, we
  use the median distance. In this way, we considered a sample of 2367 OB stars in
  this study.

\section {Distribution of $z$ with the age}
It is a well known fact that OCs are born and distributed throughout the Galactic
disk. Young clusters are normally seen in the thin disk while old clusters are found
mainly in the thick disk of the Galaxy which van den Bergh (2006) termed as a
`{\it cluster thick disk}'. In order to study the $z$ distribution of clusters with
their age, we assemble the clusters according to their $\log({\rm Age})$ in 0.2 bins dex
in width and estimate a mean value of $z$ for each bin.
%-----------------------  Fig. 2 --------------------------------------------
\begin{figure*}
\includegraphics[height=12.0cm,width=16.0cm]{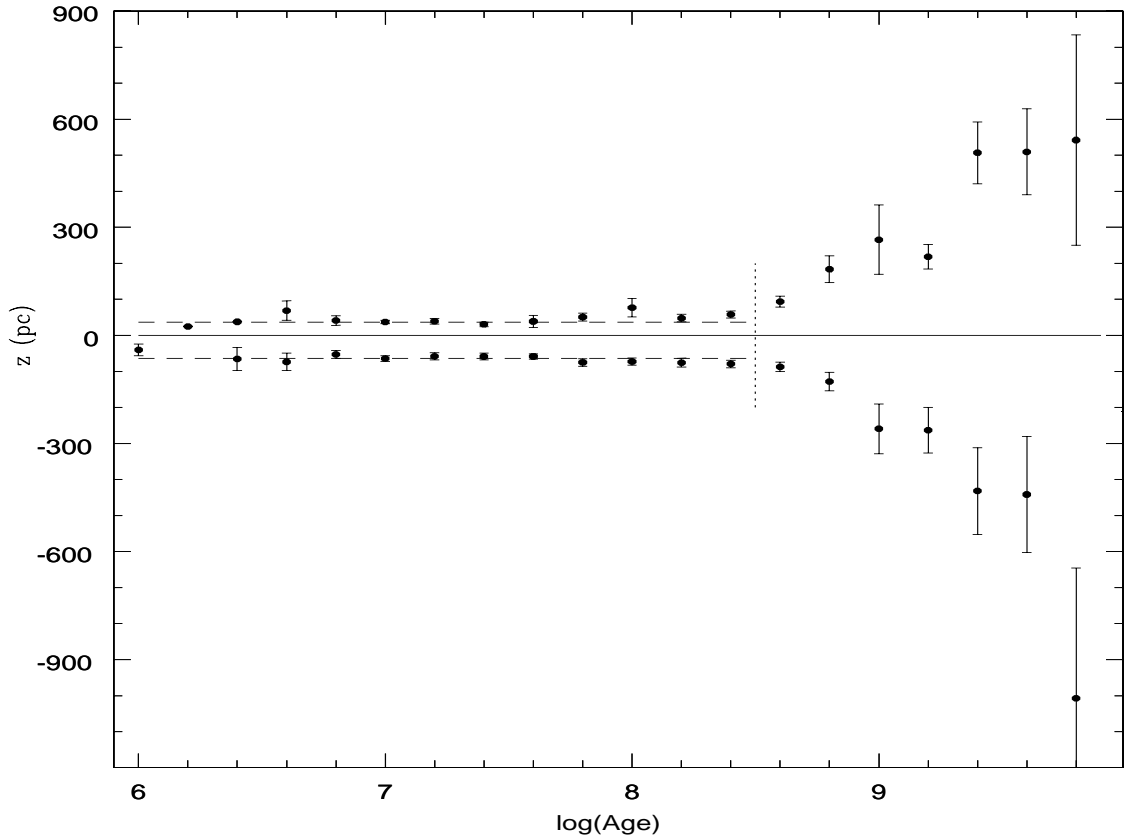}
\caption{The distribution of mean $z$ with $\log({\rm Age})$. A vertical dotted line shows
upper boundary for the age limit considered as YOCs in the present study. The
horizontal dashed lines are drawn to represent the weighted mean $z$ value of
the YOCs in the $z>0$ and $z<0$ regions. Note that there is one cluster of $\log({\rm Age})
= 10.0$ ($z \sim -172$ pc) which is not shown in the plot.}
\end{figure*}
%-------------End Fig. 2-----
A distribution of mean $z$ vs
$\log({\rm Age})$ is plotted in Fig. 2 which clearly demonstrates that the distribution of
clusters perpendicular to the GP has a strong correlation with their ages. While
clusters with $\log({\rm Age}) < 8.5$ ($\sim$ 300 Myrs) have almost a constant width
of $z$ distribution in both the directions of the GP, clusters older than this
have mean $z > 100$ pc which is continuously increases with the age. This indicates that
the thickness of the Galactic disk has not changed substantially on the time scale of
about 300 Myrs and most of the OCs, in general, formed somewhere inside $\pm$ 100 pc
of the GP. A similar study carried out by Lyng\.{a} (1982) using a smaller sample of 338
OCs found that clusters younger than one Gyr formed within $\sim$ 150 pc
of the Galactic disk. It is quite apparent from the figure that the clusters with
$\log({\rm Age}) > 8.5$ are found not only far away from the GP but are also highly scattered in their
distribution. However, this is not unexpected since it is a well known fact that
clusters close to GP gets destroyed with the time in a timescale of a few hundred
million years due to tidal interactions with the Galactic disk and the bulge, encounters
with the passing giant molecular clouds or mass loss due to stellar evolution. The few remaining
survivors reach to outer parts of the Galactic disk
(cf. Friel (1995), Bergond, Leon \& Guibert (2001)). If we just consider the clusters with
$\log({\rm Age}) < 8.5$, which we describe as YOCs in our following analysis, we find that
the 226 clusters ($\sim$ 38\%) lie above GP while 363 clusters ($\sim$ 62\%) lie below
GP. The asymmetry in cluster density above and below the GP is a clear indication of
inhomogeneous distribution of clusters around GP. This asymmetry can be interpreted
as due to the location of the Sun above the GP, displacement of the local dust layer from
the GP or asymmetry in the distribution of young star formation near the Sun with
respect to the GP or a combination of all these effects as pointed out by the van den
Bergh (2006). However, it is generally believed that it is the solar offset which
plays a major role in this asymmetry.

When we estimate weighted mean displacement along the GP for the clusters lying
within $\log({\rm Age}) < 8.5$, we find a value of $z = 37.0 \pm 3.0$ pc above the GP and 
$z = -64.3 \pm 2.9$ pc below the GP. If we consider a plane defined by the YOCs at
$z_{yoc}$, then $z_{yoc}$ can be expressed as, 
\[
z_{yoc} = \frac{n_1 z_1 + n_2 z_2}{n_1+n_2},
\]
where $z_1$ and $z_2$ are the mean $z$ for the YOCs above and below the GP respectively;
$n_1$ and $n_2$ are number of YOCs in their respective regions. This gives us a
value of $z_{yoc} = -25.4 \pm 3.0$ pc. If the observed asymmetry in the $z$ distribution of
YOCs is indeed caused by the solar offset from the GP then the negative mean displacement
of $z$ perpendicular to GP can be taken as $z_\odot$ (towards north direction) which
is about 25.4 pc. 
%------------- Fig. 3 --------------
\begin{figure*}
\includegraphics[height=16.0cm,width=14.0cm]{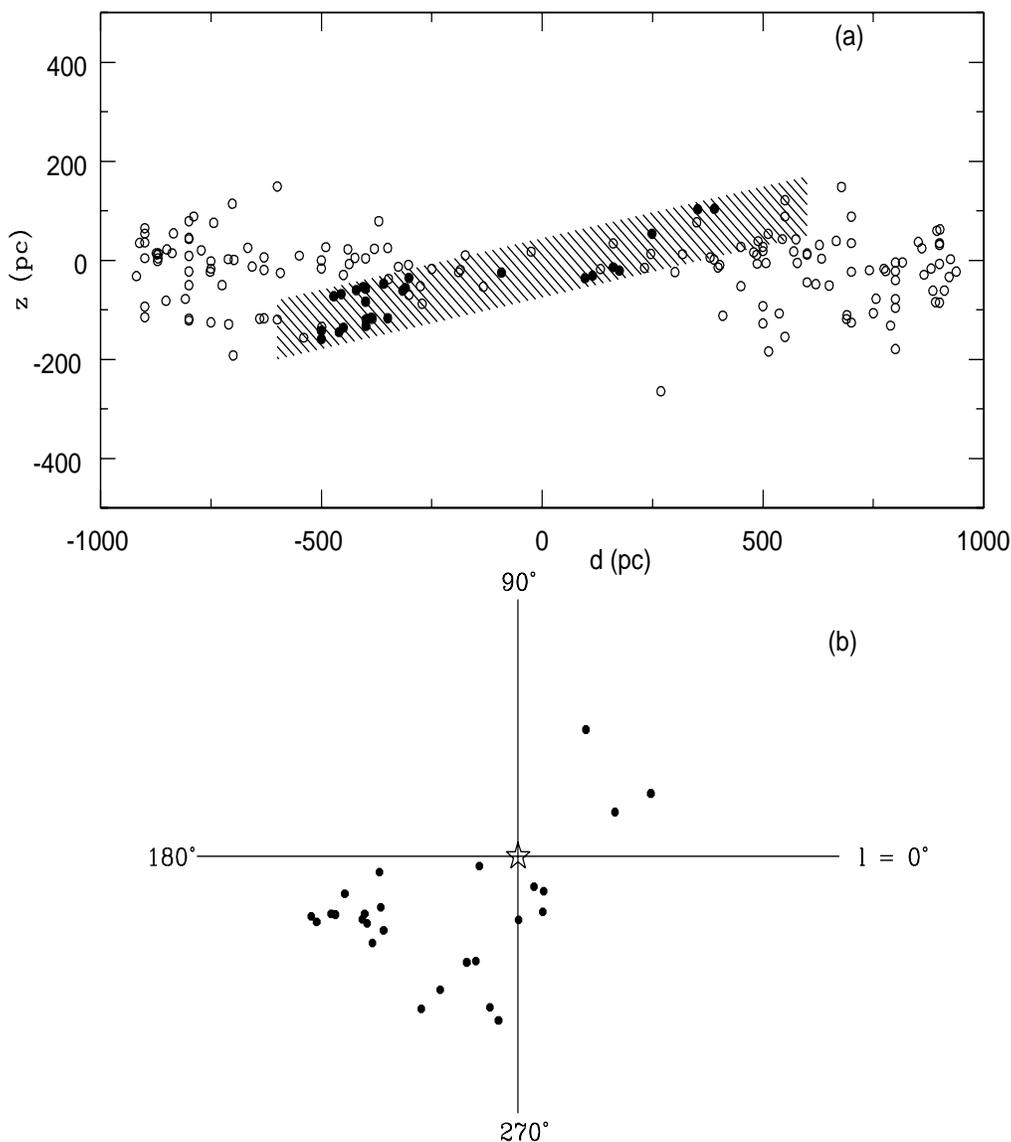}
\caption{The distribution of YOCs in the $d-z$ plane (a). Clusters towards
Galactic center direction are assigned positive distances while clusters towards
Galactic anti-center direction are assigned negative distances. Only clusters with
$|d| < 1$ kpc are plotted here for the clarity. Dark points in the shaded region
indicate the YOC's which could be associated with the GB and XY-distribution of
these 26 GB members on the GP is shown in (b) where clusters are positioned by
their distance from the Sun which is marked by a star at the center.}
\end{figure*}
%------------- End Fig. 3 --------------

However, it is a well known fact that a large fraction of the young populations
with ages under 60 Myrs in the immediate solar neighbourhood belong to the GB
(Gould 1874, Stothers \& Frogel 1974, Lindblad 1974). It is widely believed
that this belt is associated with a large structure of the interstellar matter including
reflection nebulae, dark clouds, HI gas, etc. and is tilted by about 18 deg with
respect to the GP and is stretches out to a distance of about 600 pc distance from the Sun
(Taylor, Dickman \& Scoville 1987, Franco et al. 1988, P\"{o}ppel 1997). In our sample of 589 clusters,
we found 38 such clusters which confined in the region of
600 pc from the Sun and have age below 60 Myrs. Out of the 38 clusters, 26 ($\sim 68\%$)
follow a specific pattern in the $d-z$ plane as shown by the dark points in the
shaded region of Fig. 3(a) which is slightly tilted with respect to the GP and
resembles the GB. The association of these clusters with the GB seems to be
confirmed by the fact that 23 out of 26 YOCs are clumped in the longitude range of
about 180-300 degrees as shown in Fig. 3(b). This contains the most significant
structures accounting for the expansion of the GB
(Torra, Fern\'{a}ndez \& Figueras 2000). A mean and median age of these 26 YOCs are 24.4 and 21.2 Myrs
respectively. Although no detailed study has been carried out on the fraction of
the clusters actually belonging to the GB, however, on the
basis of 37 clusters in the $\log({\rm Age}) < 7.9$ which lie within a distance
of 500 pc from the Sun, PKSS06 found that about 55\% of the clusters could be members of
the GB. On the basis of OB stars in the Hipparcos catalogue, Torra et al. (2000)
estimated that roughly 60-65\% of the stars younger than 60 Myr in the solar neighbourhood
belong to the GB. Although it is difficult to decide unambiguously which clusters belong to
the GB, we believe that most of these 26 YOCs could be associated with the GB
instead of the Local Galactic disk (hereafter LGD). Hence to reduce any systematic
effect on the determination of $z_\odot$ due to contamination of the clusters belong
to the GB, we excluded all these 26 clusters from our subsequent analysis except when otherwise
stated. When we re-derived the value of $z_\odot$ from the remaining 563 clusters, we find
it to be $22.9 \pm 3.4$ pc north of the Galactic plane. A further
discussion on the $z_\odot$ and its dependence on various physical parameters shall
be carried out below.

%------------- Fig. 4 --------------
\begin{figure*}
\includegraphics[height=16.0cm,width=16.0cm]{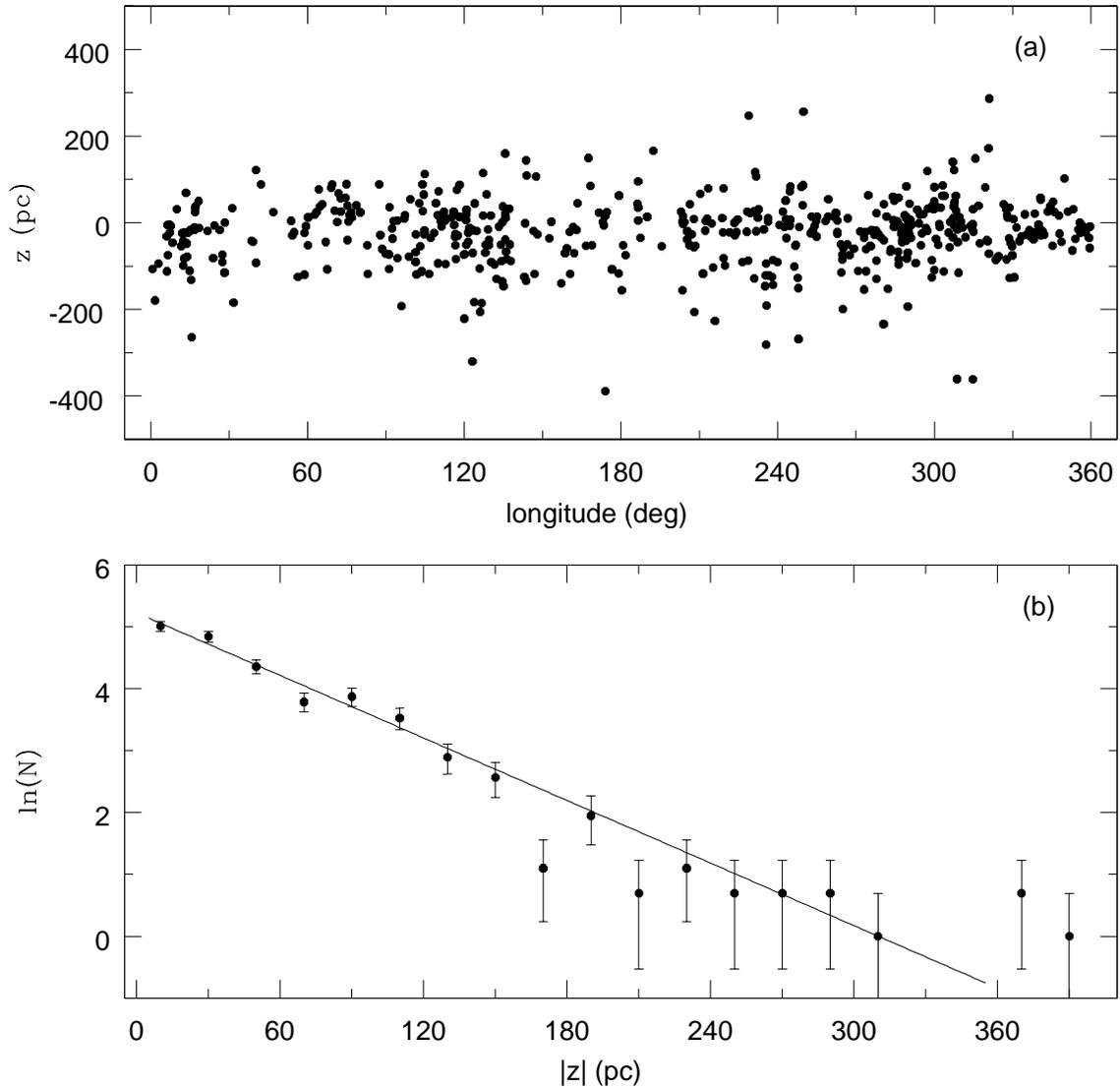}
\caption{The distribution of YOCs in the $l-z$ plane (a) and their density
distribution as a function of $z$ (b). The continuous line represents a least
square fit to the points.}
\end{figure*}
%------------- End Fig. 4 --------------

\section {Distribution of $z$ with the maximum heliocentric distance} 
\subsection {$z_\odot$ from YOCs}
Various studies indicate that the plane of symmetry defined by the OCs is inclined
with respect to the GP (Lyng\.{a} 1982, Pandey, Bhatt \& Mahra 1988, JOS05). If this is the case,
then $z_\odot$ shall be dependent on the distance of OCs from the Sun and inclination
angle between the two planes. Therefore, a simple determination of $z_\odot$ considering
all the OCs could be misleading. To examine to what extent $z_\odot$ depends on the
distance, we study the distribution of clusters and their mean displacement from the GP as 
a function of the heliocentric distance ($d_{max}$) taking advantage of the OCs observed
up to a large distance.
Since YOCs are primarily confined closer to the GP as discussed in the previous section,
it seems worthwhile to investigate $z_\odot$ using only YOCs despite the fact that the
YOCs are generally embedded in dust and gas clouds and many are not observed up to a
large distance. Although we found that some young clusters are reported as far as
9 kpc from the Sun but only less than 5\% YOCs are observed beyond 4 kpc, most of them
in the anti-center direction of the Galaxy which we do not include in our analysis.
Following all the above cuts, we retain only 537 YOCs
observed up to 4 kpc from the Sun as a working sample for the present study. Their
distribution normal to the GP as a function of Galactic longitude is plotted in Fig. 4(a).

Fig. 4(b) shows the logarithmic distribution of the YOCs as a function of $|z|$. Here we derive
the number density in bins of 20 pc and error bars shown in the y-axis is the Poisson
error. Following an exponential-decay profile, we estimate
a scale height for the YOCs as $z_h = 59.4^{+3.3}_{-3.0}$ pc which is represented by a
continuous straight line in the figure. However, a careful look in the figure suggests that the
$z_h$ could be better described by the YOCs lying within $z = \pm 250$ pc and a least square
fit in this region gives a value of $z_h = 56.9^{+3.8}_{-3.4}$ pc. 

It is however interesting to see if the scale height shows any shift in its value when considering
a possible displacement of the cluster plane from the GP. In order to analyse any effect of the
displacement on $z_h$, we shift the cluster plane by 10, 15, 20 and 25 pc from the GP and
recalculate $z_h$ using YOCs within $z < 250$ pc. Our results are given in Table 1. It is
seen that these values of $z_h$ are quite consistent and we conclude that the solar offset
has no bearing in the determination of scale height. Using a sample of 72 OCs younger than
800 Myrs, Janes \& Phelps (1994) reported a scale height of $z_h \sim 55$ pc. 
Recently Bonatto et al. (2006) derived a scale height of $z_h = 48 \pm 3$ pc using a sample
of clusters younger than 200 Myrs, however, they have also found a larger $z_h$ when considering
OCs older than 200 Myrs. PKSS06 obtained a scale height of $z_h = 56 \pm 3$ pc
using the OCs within 850 pc from the Sun. Our value of $z_h = 56.9^{+3.8}_{-3.4}$ pc obtained
with the YOCs within 4 kpc from the Sun and $z < 250$ pc is thus consistent with these
determinations.
%----------------------------- Table 1 -----------------------------
\begin{table}
\centering
\caption{Scale heights determined due to various offsets between cluster plane and GP. All
the values are in pc.}
\begin{tabular}{cc} \hline
\vspace{0.1cm}
shift       &    $z_h$              \\  \hline
  0         & $56.9^{+3.8}_{-3.4}$  \\ \\
 10         & $55.1^{+3.3}_{-2.9}$  \\ \\
 15         & $54.7^{+3.2}_{-2.9}$  \\ \\ 
 20         & $57.2^{+3.9}_{-3.5}$  \\ \\
 25         & $56.6^{+3.9}_{-3.3}$  \\ \\
\hline
\end{tabular}
\end{table}
%-----------------------------------------------------------------

An important issue that needs to be addressed in the determination of $z_\odot$ is the possible
contamination by the outliers which are the objects lying quite far away from the GP that can
seriously affect the $z_\odot$ estimation. Hence it is worthwhile at this point to
investigate $z_\odot$ using a subsample of YOCs in different $z$ zone excluding the
clusters far away from the GP without significantly reducing the number of clusters.
If the observed asymmetry in the cluster distribution is really caused by an offset of the Sun
from the GP, then a single value of $z$ should result from the analysis. In order to study
$z_\odot$ distribution using YOCs, we select three different zones normal to the $z = 0$ plane
considering the clusters within $|z| < 150$ pc, $|z| < 200$ pc and $|z| < 300$ pc. Here, we
have not made smaller zones than $|z| = 150$ pc keeping in mind the fact that accounting
lesser number of YOCs could have resulted in a larger statistical error while zone larger
than $|z| = 300$ pc can cause significant fluctuations due to few but random clusters observed
far away from the GP.
%------------- Fig. 5 --------------
\begin{figure*}
\includegraphics[height=12.0cm,width=16.0cm]{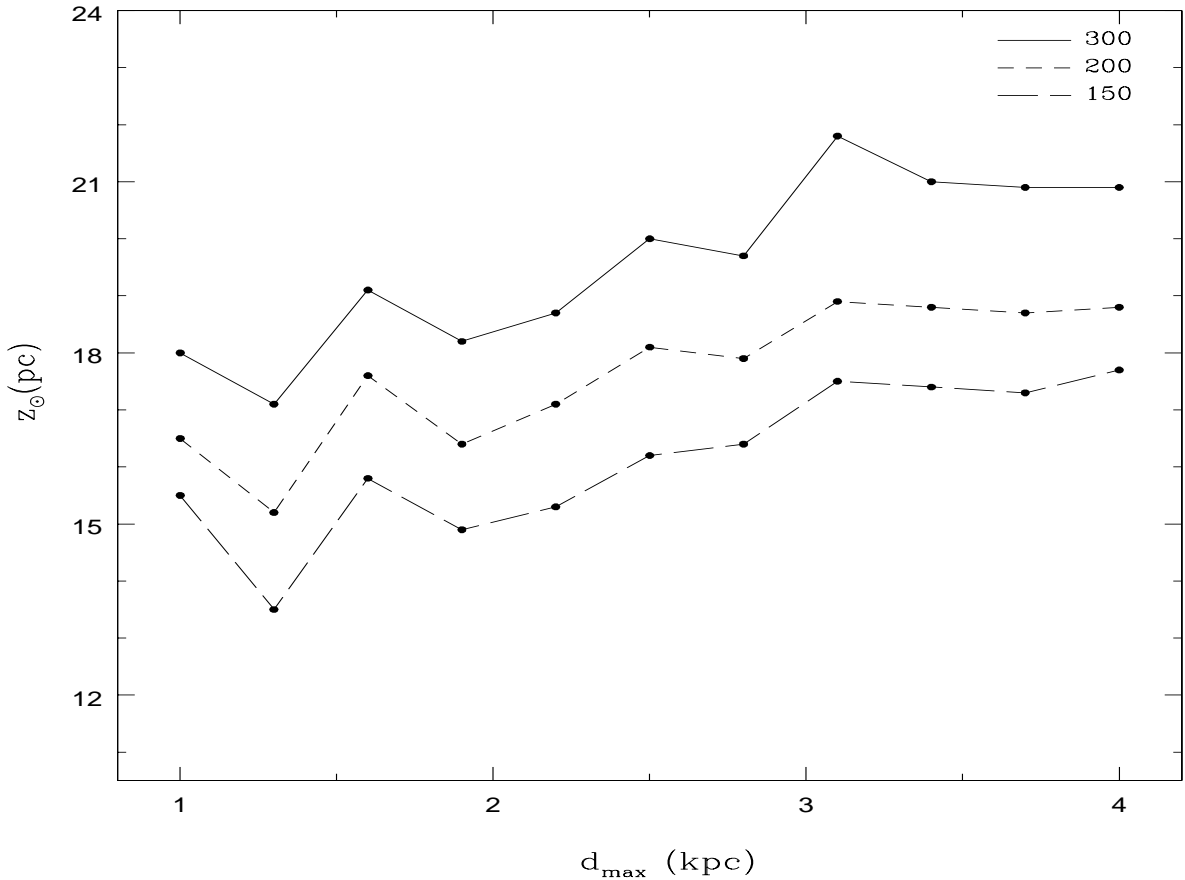}
\caption{The variation in $z_\odot$ with the maximum distance of YOCs from the Sun
(see text for the detail).}
\end{figure*}
%------------- End Fig. 5 --------------
To determine $z_\odot$, we keep on moving the mid-plane towards the southwards direction in bins of
0.1 pc to estimate the mean $z$ till we get the mean value close to zero i.e. a plane
defined by the YOCs around which the mean $z$ is zero within the given zone that is
in fact equivalent to $z_\odot$. This approach of a running shift of $z$ in order to
determine $z_\odot$ is preferred over the simple mean to remove any biases owing to
the displacement of the cluster plane itself towards the southwards direction. Hence it
gives a more realistic value of the $z_\odot$. We estimate $z_\odot$ with different
cut-off limits in $d_{max}$ using an increment of 0.3 kpc in each step and for all the
three zones. The variation in $z_\odot$ with $d_{max}$ for all the zones is illustrated in
Fig. 5. The figure gives a broad idea
of the variation in $z_\odot$ which increases with the increasing distance as well
as zone size, however, it has to be noted that the range of variation is very
small and varies between $\sim$ 13 to 21 pc throughout the regions. 

Here, it is necessary to look into the increasing trend in $z_\odot$ whether it is internal
variation or due to our observational limitations. We note that 21 out of 25 YOCs observed
beyond 1 kpc in the region $|z| > 150$ pc are
observed in the direction of $l = 120^\circ<l<300^\circ$. Moreover, most of these young
clusters are observed below GP and majority of them are located in the direction of
$l \sim 200^\circ<l<300^\circ$. This could be due to low interstellar extinction in the
Galactic anti-center direction which is least around the longitude range
$220^\circ - 250^\circ$ (Neckel \& Kare 1980, Arenou, Grenon \& G\'{o}mez 1992,
Chen et al. 1998). Based on the
study of extinction towards open clusters from the same catalogue of Dias et al.
(2002), we found the direction of minimum extinction towards $l \sim 230^\circ$
below the GP (JOS05).
Hence a lower extinction allows us to have a higher observed cluster density in the
surrounding area of the $l \sim 230^\circ$ as well as observable up to farther distance
which reflected in our larger value of $z_\odot$ with the increase of the distance.
Therefore, we conclude that the larger $z_\odot$ values
obtained with the bigger zone or greater distance is not due to internal
variation in $z_\odot$ but due to our observational constraint. In general, we found
a value of $17\pm3$ pc for the $z_\odot$.

%------------- Fig. 6 --------------
\begin{figure*}
\includegraphics[height=16cm,width=16cm]{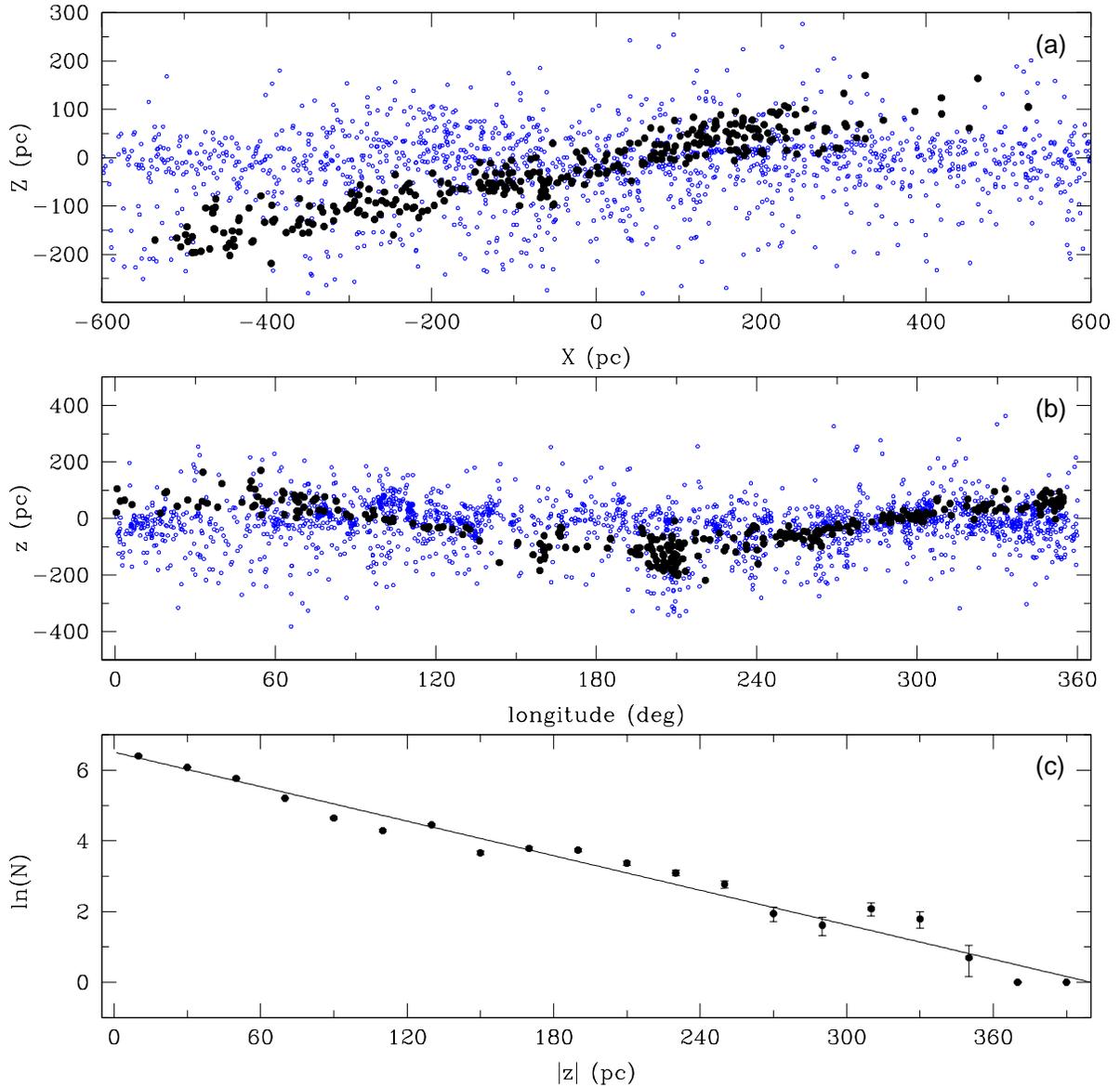}
\caption{The X-Z distribution of the OB stars in (a). The open circles represent the
OB stars belong to LGD and filled circles represent possible GB members. The x-axis
is drawn for only $\pm 600$ pc to show the GB members clearly which is quite evident
in the diagram. Their distribution in the $l-z$ plane is drawn in (b). A number density
distribution of the OB stars belong to the LGD as a function of $z$ is
shown in (c). The continuous line here indicates a least square fit to the points.}
\end{figure*}
%------------- End Fig. 6 --------------
\subsection {$z_\odot$ from OB stars}
Since YOCs are on an average more luminous than the older clusters and also possess a
large number of OB stars hence lends us an opportunity to compare the results with
the independent study using massive OB stars which are also a
younger class of objects and confined very close to the GP. In the present analysis, we
use 2367 OB stars which are strongly concentrated towards the GP as those of the YOCs.
However, a natural problem in the determination of $z_\odot$ is to separate the
OB stars belonging to the GB with the LGD. The issue has already been dealt with
a great detail by several authors (Taylor, Dickman \& Scoville 1987, Comeron,
Torra \& Gomez 1994,
Cabrera-Ca\~{n}o, Elias \& Alfaro 1999, Torra, Fern\'{a}ndez \& Figueras 2000). A recent
model proposed by the ECA06 based on the three dimensional
classification scheme allows us to determine the probability of a star belonging to the
GB plane or LGD. A detailed discussion of the method can be found in the ECA06 and we
do not repeat it here.
Though it is not possible to unambiguously classify the membership of the stars among
two populations but to statistically isolate the GB members from our sample, we used the
results derived for the GB plane by the ECA06 through the exponential probability density
function for the O-B6 stars selected from the Hipparcos catalogue while we
used an initial guess value of 60 pc and -20 pc for the scale height and $z_\odot$ respectively
for the GP. Since typical maximum radius of the GB stars is not greater than about 600 pc
(Westin 1985, Comeron, Torra \& Gomez 1994, Torra, Fern\'{a}ndez \& Figueras 2000), we
search OB stars belonging to GB up to this distance only. 

Following the ECA06 method, we found that 315 stars out of 2367 OB stars of our data sample
belong to the GB. Further, 22 stars do not seem to be associated with either of the planes.
In this way, we isolate 2030 OB stars belonging to the LGD which are used in our
following analysis. A $X-Z$ distribution of the OB stars is shown in Fig. 6(a) (in the
Cartesian Galactic coordinate system, positive $X$ represents the axes pointing to the
Galactic center and positive $Z$ to the north Galactic pole) and their distribution in
the GP as a function of Galactic longitude is displayed in Fig. 6(b). A clear separation
of the GB plane from the GP can be seen in the figure which follows a sinusoidal variation
along the Galactic longitude and reaches its lower latitude at $l = 200-220^\circ$.
A number density in the logarithmic scale of the OB stars belonging to LGD is shown
in Fig 6(c) as a function of $|z|$ where stars are counted in the bins of 20 pc.
We derive a scale height of $z_h = 61.4^{+2.7}_{-2.4}$ pc from
the least square fit that is drawn by a continuous straight line in the same figure.
Ma\'{i}z-Apell\'{a}niz (2001) using a Gaussian disk model determined a value of
$z_h = 62.8 \pm 6.4$ pc which is well in agreement with our result. However, Reed
(2000) derived a broad range of $z_h \sim 25 - 65$ pc using O-B2 stars while ECA06
estimates smaller value of $34 \pm 3$ pc using O-B6 stars which are more in agreement
with the $34.2 \pm 3.3$ pc derived with the self-gravitating isothermal disk model of
Ma\'{i}z-Apell\'{a}niz (2001).
%------------- Fig. 7 --------------
\begin{figure*}
\includegraphics[height=12.0cm,width=16.0cm]{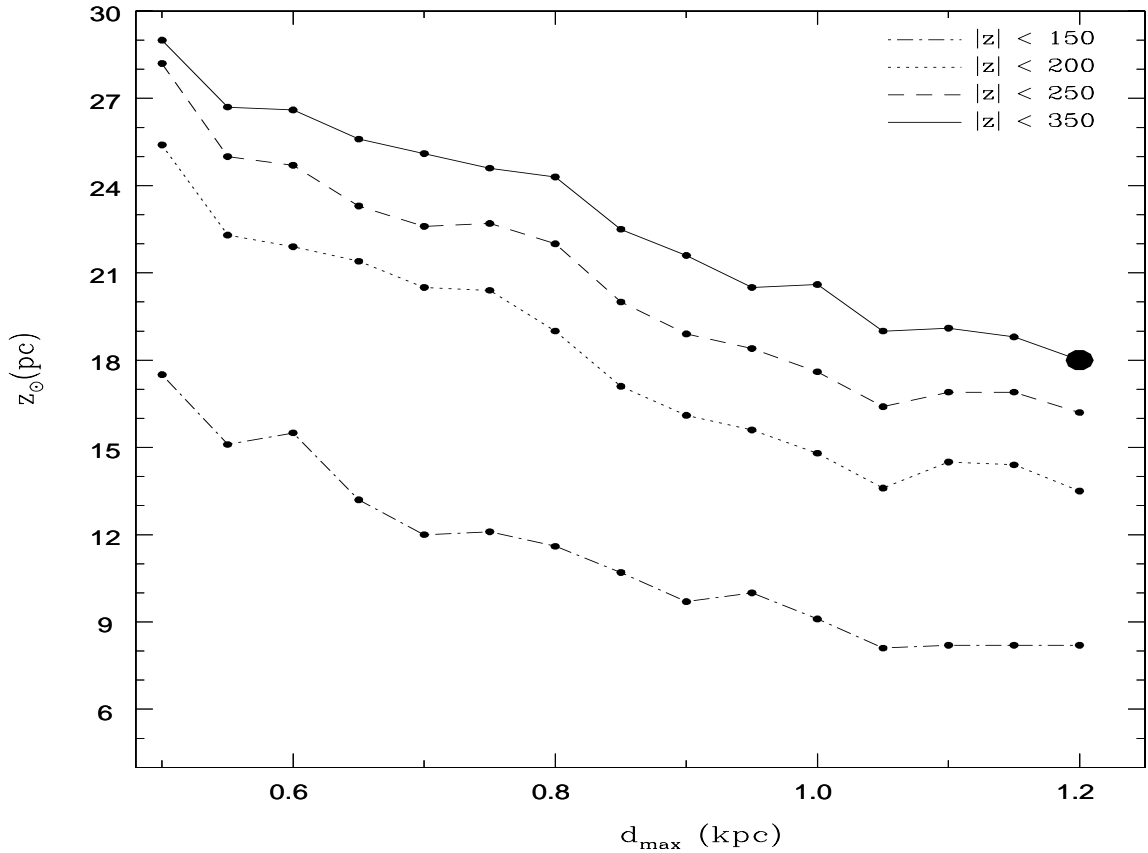}
\caption{A similar plots as in Fig. 5 but for the OB stars. A big dot here
represents the $z_\odot$ using all the OB stars considered in our study.}
\end{figure*}
%------------- End Fig. 7 --------------

It is seen in Fig. 6(b) that the OB stars are sparsely populated around the GP in
comparison of the YOCs and a significant fraction of them are below $z = -150$ pc. In
order to study the $z_\odot$ distribution with $d_{max}$, we here make four different
zones normal to the $z = 0$ plane considering the OB stars within $|z| < 150$ pc, $|z| < 200$
pc, $|z| < 250$ and $|z| < 350$ pc. The $z_\odot$ is estimated by the same
procedure as followed for the YOCs. A variation in the $z_\odot$ with $d_{max}$
is illustrated in Fig. 7 where we have made a bin size of 50 pc. It is seen that
$z_\odot$ derived in this way for the OB stars
show a continuous decay with the $d_{max}$ as well as size of the zone which seems to
be due to the preferential distribution of the OB stars below the GP. When we draw the
spatial distribution of OB stars in the X-Y coordinate system in Fig. 8, we
notice that most of the OB stars are not distributed randomly but concentrated in the
loose group of the OB associations. This difference in density distribution of OB stars
could be primarily related with the star forming regions.
The number of OB stars below the GP are always found to be greater than the OB stars
above the GP in all the distance bin of 100 pc. However, in the immediate solar
neighbourhood within 500 pc distance, OB stars below the GP are as 
much as twice than those above the GP. 
%Most of these OB stars are distantly located normal to
%the GP which are indicated by the larger size of symbols in Fig. 8.
This is clearly
a reason behind a large value of $z_\odot$ in the smaller $d_{max}$ value which
systematically decreases as more and more distant OB stars are included. 
A mean value of $19.5 \pm 2.2$ pc was obtained by Reed
(2006) using the same catalogue of 2397 OB stars, albeit without removing the GB members.
In fact this is also noticeable in the present study (see big dot in Fig. 7). However,
we cannot give a fixed value of $z_\odot$ from the present analysis of the OB stars
as it depends strongly on the $d_{max}$ as well as selection of the $z$ cut-off.

%------------- Fig. 8 --------------
\begin{figure*}
\includegraphics[height=14.0cm,width=14.0cm]{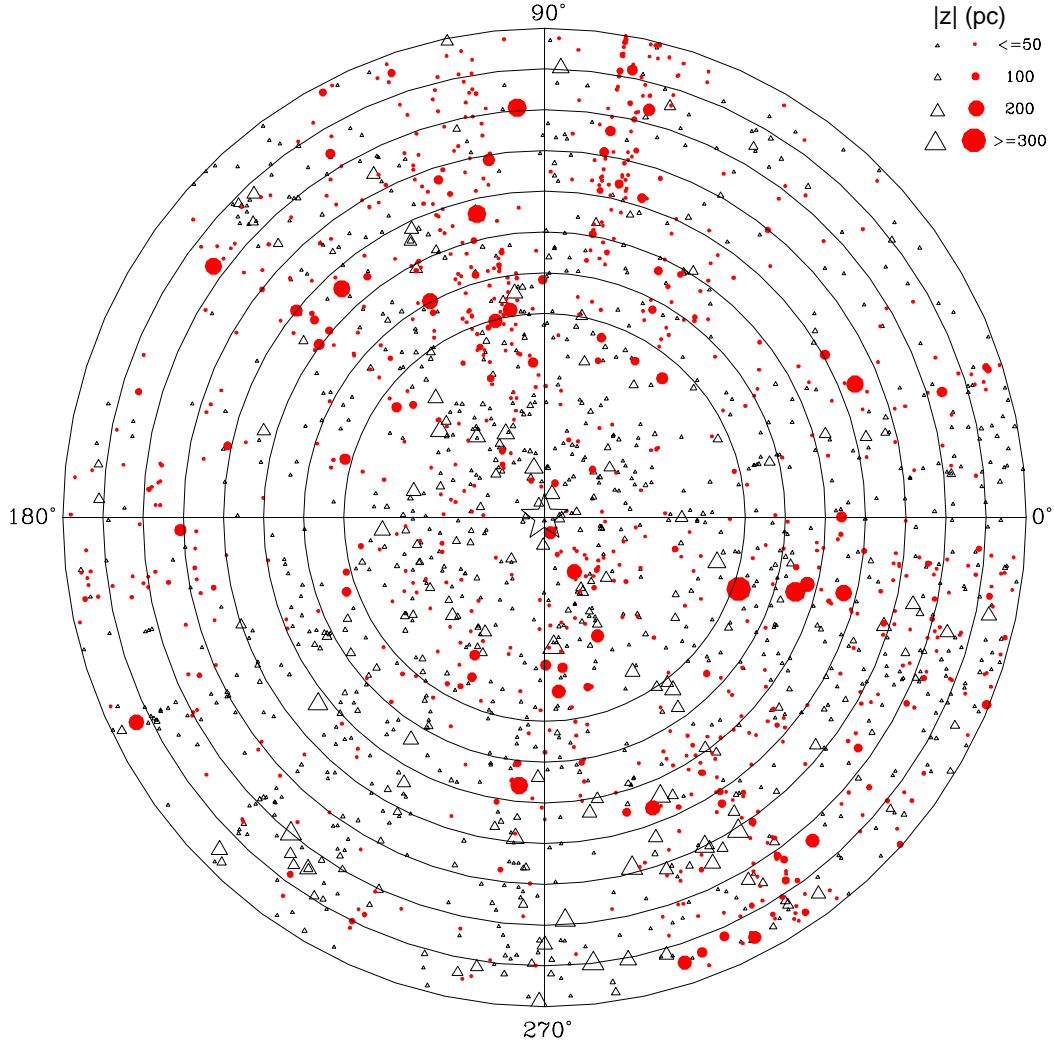} 
\caption{A spatial distribution of the OB stars belonging to the LGD projected on the
GP where position of the Sun is shown by a star symbol at the center. Open triangles
and filled circles represent the stars below and above the GP respectively. Size of
the points signify the distance of OB stars normal to the GP as indicated at the top of
the diagram. The co-centric circles at an equal distance of 100 pc from 500 pc to
1200 pc are also drawn.}
\end{figure*}
%------------- End Fig. 8 --------------

\section {Exponential decay of the $z$ distribution}
It is normally assumed that the cluster density distribution perpendicular to
the GP could be well described in the form of a decaying exponential away from the GP,
as given by,
\[
N = N_0 exp\left[-\frac{|z+z_\odot|}{z_h}\right] ,
\]
%------------- Fig. 9 --------------
\begin{figure*}
\includegraphics[height=12.0cm,width=16.0cm]{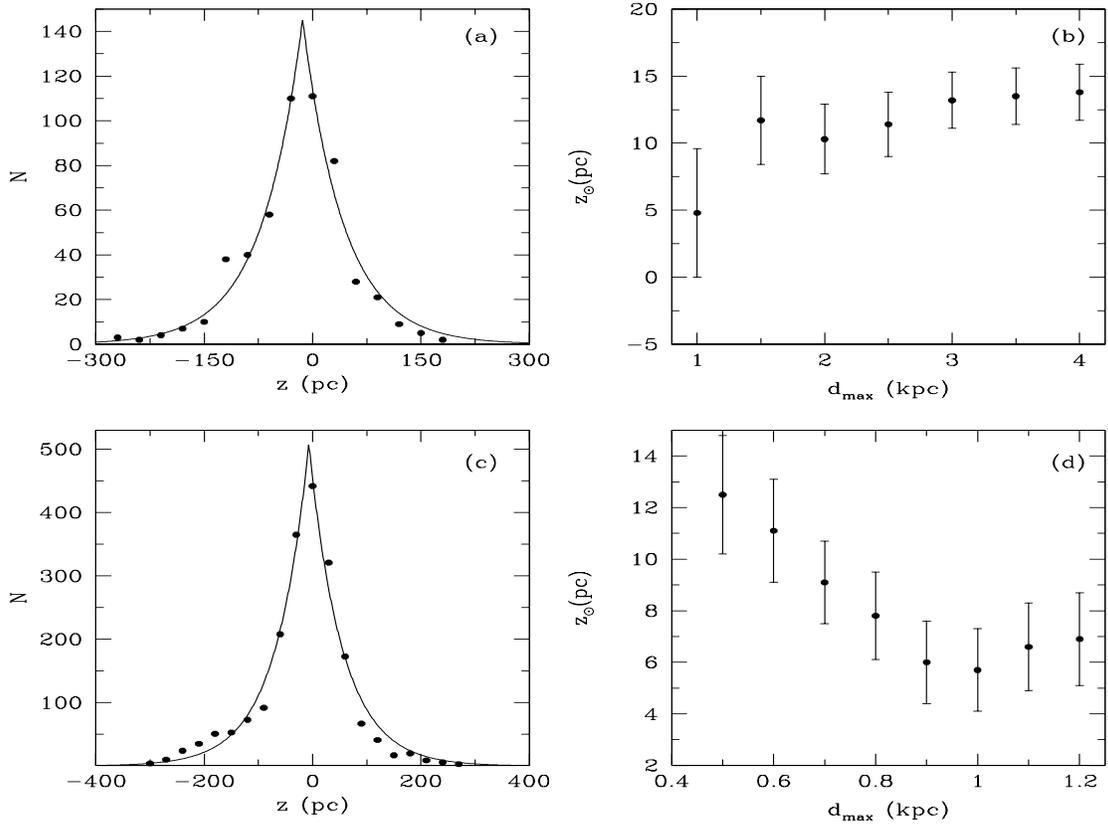}
\caption{The $z$ distribution for all the OCS within $|z| < 300$ pc and $d < 4$ kpc
(a). A least square exponential decay profile fit is also drawn by the continuous
line. The $z_\odot$ derived from the fits for different $d_{max}$ is shown in (b).
The same is shown for the OB stars in (c) and (d).}
\end{figure*}
%------------- End Fig. 9 --------------
where $z_\odot$ and $z_h$ are the solar offset and scale height respectively.
We determine $z_\odot$ by fitting the above function. For example in Fig. 9(a),
we have drawn $z$ distribution in 30 pc bin considering all the 537 YOCs which
lie within $|z| < 300$ pc and $d < 4$ kpc. Since we have already derived the scale height
for the YOCs as 56.9 pc in our earlier section hence kept it fixed in the present fit.
A least square exponential is fitted for all the distance limits. Here we do not
divide the data sample in different zones of $z$ as we have done in the previous section
since only the central region of $\pm$ 150 pc has significant effect on the
determination of solar offset in the exponential decay method as can be seen
in Fig. 9(a).

Our results are shown in Fig. 9(b) where we have displayed $z_\odot$ derived for the
YOCs as a function
of $d_{max}$. We can see a consistent value of about 13 pc for $z_\odot$ except
when only YOCs closer to 1 kpc from the Sun are considered. This may be due to
undersampling of the data in that region. Our estimate is close to the Bonatto
et al. (2006)
who reported a value of $14.2 \pm 2.3$ pc following the same approach, however,
clearly lower in comparison of $z_\odot$ determined in the previous section.
Here, it is worth to point out that following the same approach PKSS06 found a
significantly large value of $z_\odot$ ($\sim 28-39 \pm 9$ pc) when considering
only those clusters within $\log({\rm Age}) < 8.3$. However, the value of
$z_\odot$ substantially comes down to $8\pm8$ pc for the clusters in the age
range of $8.3 < \log({\rm Age}) < 8.6$ in their study. If we confine our sample to
$\log({\rm Age}) < 8.3$ only, we find that $z_\odot$ increases marginally up to
$14.6$ pc which is not quite different than our earlier estimate but
still considerably lower than the PKSS06 and we suspect that their values are
overestimated by a significant factor.

A similar study for the $z$ distribution of OB stars is also carried out and our
results are shown in Fig. 9(c), as an example, considering all the
data sample. The resultant variation of $z_\odot$ for the different $d_{max}$
are shown in Fig. 9(d). It is clearly visible that $z_\odot$ varies in the range of
6 to 12 pc which is substantially lower in comparison of the values obtained in the
previous method for the same data set. Reed (1997, 2000) also reported a similar
lower value of $\sim$ 6 to 13 pc for the
$z_\odot$ using exponential model. A significant feature we notice here is that
the $z$ distribution to the left and right of the peak do not seem symmetric particularly
in the bottom half of the region where exponential fit in the $z > z(N_{max})$ region is
higher than their observed value while reverse is the case for the $z < z(N_{max})$ region.
Therefore, a single exponential profile fit to the distribution of the OB stars for the
whole range results in a large $\chi^2$ since points are well fitted only over a short
distance interval around the mid-plane. This may actually shift $z_\odot$ towards the lower
value which results in an underestimation for the $z_\odot$ determination. We believe that
a single value of $z_\odot$ determined through exponential decay method is underestimated
and needs further investigation.

\section {Distribution of $z$ with the Galactic longitude}
A distribution of clusters in the Galactic longitude also depends upon the Age
(Dias \& L\'{e}pine, 2005) and it is a well known fact that the vertical displacement
of the clusters from the GP is correlated with the age of the clusters. Hence, one
alternative way to ascertain the mean displacement of Sun from the GP is to study
the distribution of YOCs and OB stars projected on the GP as a function of the Galactic
longitude where it is noticeable that the distribution follows an approximately
sinusoidal variation. We estimated $z_\odot$ in this way in our earlier study
(JOS05) although analysis there was based on the differential distribution of
interstellar extinction in the direction of OCs.

%------------- Fig. 10 --------------
\begin{figure*}
\includegraphics[height=12.0cm,width=16.0cm]{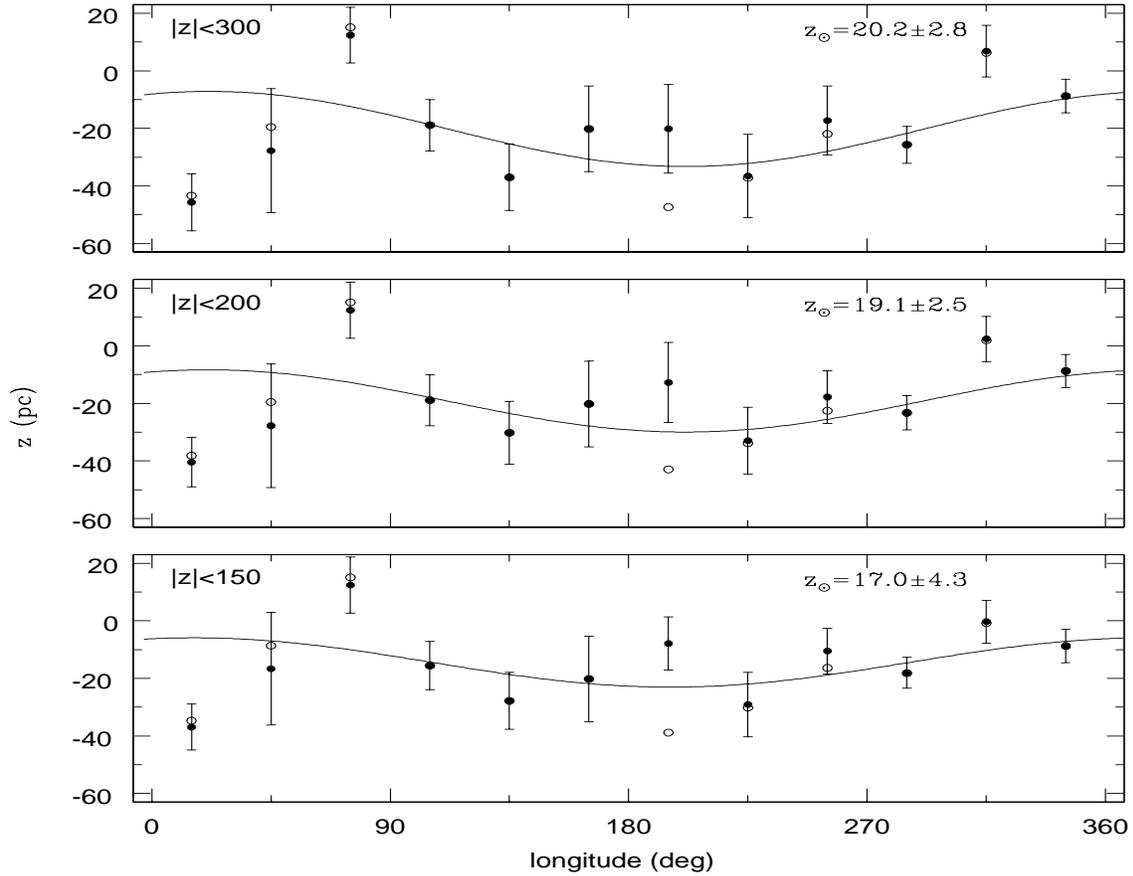}
\caption{Mean $z$ of the YOCs as a function of Galactic longitude. Here open and
filled circles represent the $z$ distribution with and without GB members
respectively. A least squares sinusoidal fit is drawn by the continuous line.
Respective regions in $|z|$ and $z_\odot$ determined from the fit are shown at the
top of each plot.}
\end{figure*}
%------------- End Fig. 10 --------------
To study the variation of $z$ as a function of Galactic longitude, we assemble YOCs
in $30^\circ$ intervals of the Galactic longitude and mean $z$ is determined for
each interval. Here we again divide the YOCs in three different zones as discussed in
Sect. 4 and the results are illustrated in Fig. 10 where points are drawn by the
filled circles. Considering the scattering and error bars in mind, we do not see any
systematic trend in the $z$ variation and a constant value of $14.5\pm2.2,
17.4\pm2.6, 18.5\pm2.9$ pc (in negative direction) are found for $|z|<150$, $|z|<200$
and $|z|<300$ pc respectively. However, when we consider all the YOCs including
possible GB members as drawn by open circles in the same figure, we found a
weak sinusoidal variation as plotted in Fig. 10 by the continuous lines and has a
striking resemblance
with $z$ distribution at maximum Galactic absorption versus longitude diagram
(Fig. 8 of JOS05). We fit a function,
\[
z = -z_\odot + a sin(l+\phi) ,
\]
to the $z(l)$ distribution with $z_\odot$ estimated from the least square fits in all
the three zones and resultant values are given at the top of each panel in Fig. 10. It
is clearly visible that the $z_\odot$ estimated in this way varies between
17 to 20 pc and it is not too different for the case when GB members are excluded.
The largest shift in the mean $z$ below the GP occurs at about
$210^\circ$ which is the region associated with the GB (see Fig. 6(b)) as can be
seen by the maximum shift between filled and open circular points in Fig. 10.
%------------- Fig. 11 -----------------------------
\begin{figure*}
\includegraphics[height=12.0cm,width=16.0cm]{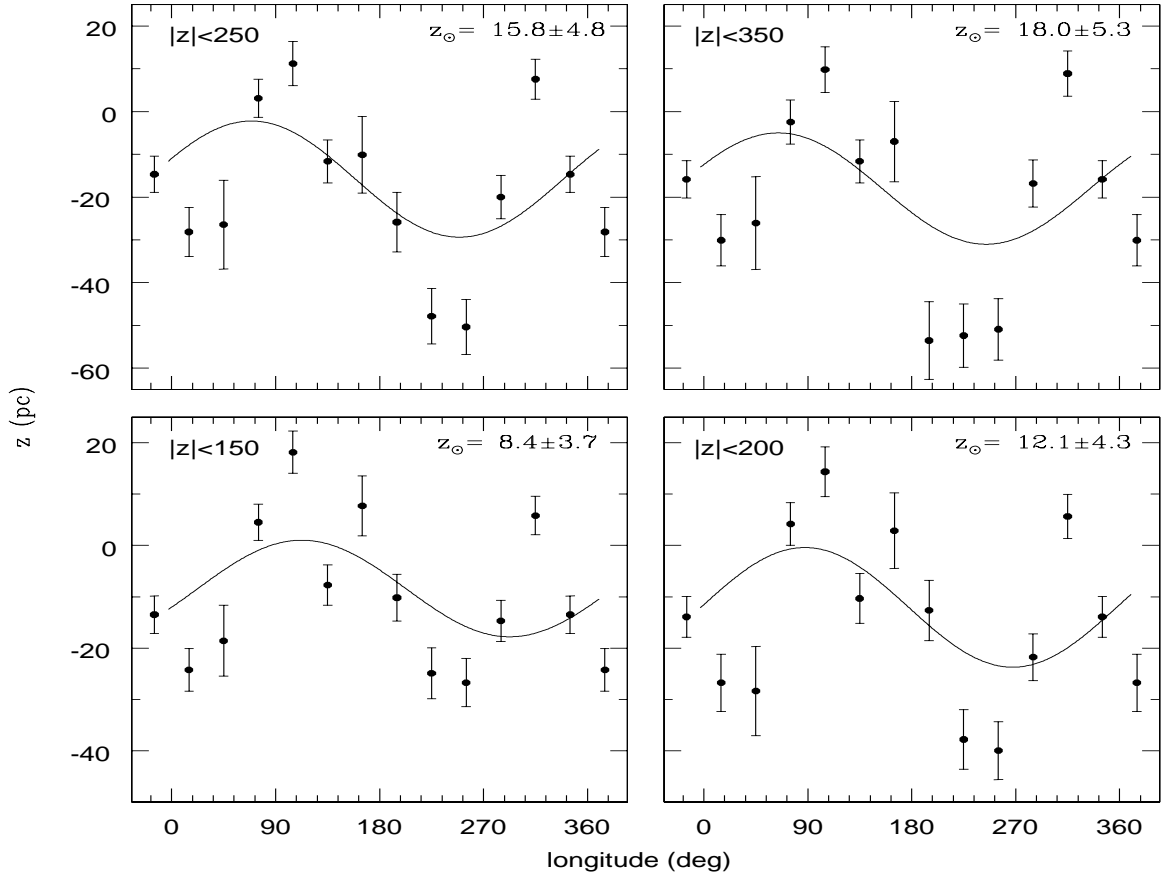}
\caption{A similar plots as in Fig. 10 but for the OB stars.}
\end{figure*}
%-------------End Fig. 11 --------------------------

In Fig. 11, we plot a similar variation for the OB stars in four different zones as
selected in Sect. 4 and it is noticeable that the sinusoidal variation is more
promising for the OB stars. The values of $z_\odot$ ranges from 8.4 to 18.0 and like
in all our previous methods, it shows a significant variation among
different $d_{max}$ for the OB stars.

It is interesting to note that mean $z$ shows a lower value in the vicinity
of $l \sim 15^\circ-45^\circ$ region in both the YOCs and OB stars. Pandey, Bhatt
\& Mahra (1988) argued that since the maximum absorption occurs in the direction of
$l \sim 50^\circ$ as well as reddening plane is at the maximum distance from the
GP in the same direction of the Galactic longitude, it may cause a lower
detection of the objects. We also found a similar result in JOS05. In his diagram
of the distribution of OCs as a function of longitude, van den Bergh (2006) also
noticed that the most minimum number of OCs among various dips lies in the region of
$l \sim 50^\circ$ where there is an active star forming region, Sagitta. However, the
lack of visible OCs are compensated by the large number of embedded clusters
detected from the 2MASS data (Bica, Dutra \& Soares 2003). We therefore attribute an
apparent dip in $z_\odot$ around the region $l \sim 50^\circ$ to the observational
selection effects associated due to star forming molecular clouds which may result
in the non-detection of many potential YOCs towards far-off directions normal to the GP.

\section {Concluding remarks}
The spatial distribution of the young stars and star clusters have been widely used to
probe the Galactic structure due to their enormous luminosity and preferential location
near the GP and displacement of the Sun above GP is one issue that has been addressed
before by many authors. In the present paper we considered a sample of 1013 OCs and 2397 OB
stars  which are available in the web archive. Their $z$ distribution
around the GP along with the asymmetry in their displacement normal to the GP allowed us
to statistically examine the value of $z_\odot$. The cut-off limit of 300 Myrs in
the age for YOCs has been chosen on the basis of their distribution in the $z-\log({\rm Age})$
plane. We have made an attempt to separate out the OCs and OB stars belonging to the
GB from the LGD. 

In our study, we have attempted three different approaches to
estimate $z_\odot$ using 537 YOCs lying within 4 kpc from the Sun. We have
studied $z_\odot$ variation with the maximum heliocentric distance and found
that $z_\odot$ shows a systematic increase when plotted as a function of $d_{max}$,
however, we noticed that it is more related to observational limitations due to
Galactic absorption rather that a real variation. After analysing these YOCs, we conclude that
$17\pm3$ pc is the best estimate for the $z_\odot$. A similar value has been obtained
when we determined $z_\odot$ through the $z$ distribution of YOCs as a function of
Galactic longitude, however, a smaller value of about 13 pc is resulted through exponential
decay method. Considering the YOCs within $z < 250$ pc, we
determined that the clusters are distributed on the GP with a scale height of
$z_h = 56.9^{+3.8}_{-3.4}$ pc and noticed that the $z_\odot$ has no bearing
in the estimation of $z_h$. A scale height of $z_h = 61.4^{+2.7}_{-2.4}$ pc
has also been obtained for the OB stars belonging to the LGD.

A comparative study for the determination of $z_\odot$ has been made using the 2030
OB stars lying within a distance of 1200 pc from the Sun and belonging to the LGD. It is
seen that the $z_\odot$ obtained through OB stars shows a substantial variation from about
8 to 28 pc and strongly dependent on the $d_{max}$ as well as
$z$ cut-off limit. It is further noted that $z_\odot$ estimated through exponential
decay method for the OB stars gives a small value in comparison of the YOCs and ranges
from 6-12 pc.
Therefore, a clear cut value of $z_\odot$ based on the OB stars cannot
be given from the present study, however, we do expect that a detailed study of OB
associations in the solar neighbourhood by the future GAIA mission may provide improved
quality and quantity of data to precisely determine $z_\odot$ in order to
understand the Galactic structure.
This paper presents our attempt to study the variation in $z_\odot$ due to selection
of the data and method of determination using a uniform sample of YOCs and OB stars
as a tool. It is quite clear from our study that the differences in approach and
choice of the data sample account for most of the disagreements among $z_\odot$ values.

\section*{Acknowledgments}
This publication makes use of the catalog given by W. S. Dias for the OCs and by
B. C. Reed for the OB stars. Author is thankful to the anonymous referee for his/her
comments and suggestions leading to the significantly improvement of this paper. The
critical remarks by John Eldridge are gratefully acknowledged.

\end{document}